\documentclass{article}

\usepackage{arxiv}

\usepackage[utf8]{inputenc} % allow utf-8 input
\usepackage[T1]{fontenc}    % use 8-bit T1 fonts
\usepackage{graphicx}
\usepackage{graphics}
\usepackage{epsfig}
\usepackage{epstopdf}
\usepackage{float}
\usepackage{color}
\usepackage{hyperref}       % hyperlinks
\usepackage{url}            % simple URL typesetting
\usepackage{booktabs}       % professional-quality tables
\usepackage{amsfonts}       % blackboard math symbols
\usepackage{nicefrac}       % compact symbols for 1/2, etc.
\usepackage{microtype}      % microtypography
\usepackage{lipsum}		% Can be removed after putting your text content
\usepackage{graphicx}
\usepackage{natbib}
\usepackage{doi}
\usepackage{overpic}
\usepackage{tikz}
\usepackage{multirow}
\usepackage{hyperref}
\usepackage{xcolor}
\usepackage[normalem]{ulem}
\usepackage{diagbox}
\usepackage[cmex10]{amsmath}
\usepackage{latexsym,amsfonts}
\usepackage{amsthm}

\graphicspath{{./img/}}

\newcommand\Mat[1]{\mathbf{#1}}
\newcommand\Vect[1]{\mathbf{#1}}
\title{Snapshot HDR Video Construction Using Coded Mask}

\author{ \href{https://orcid.org/0000-0002-7866-3117}{\includegraphics[scale=0.06]{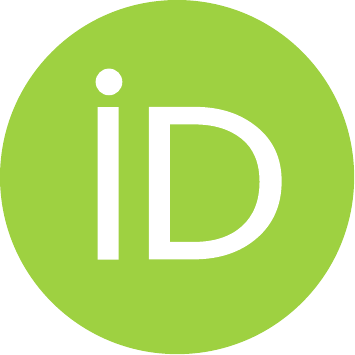}\hspace{1mm}Masheal M. Alghamdi} \\
	National Center for Data Analytics and Artificial Intelligence\\
    KACST \\
	\texttt{malghamdi@kacst.edu.sa} 
	\And
	\href{ https://orcid.org/0000-0001-6395-8521}{\includegraphics[scale=0.06]{orcid.pdf}\hspace{1mm}  Qiang Fu}\\
	Visual Computing Center\\
	KAUST\\
	Thuwal, SA\\ 
	\texttt{qiang.fu@kaust.edu.sa} \\
	\And
	\href{https://orcid.org/0000-0001-7513-0748}{\includegraphics[scale=0.06]{orcid.pdf}\hspace{1mm}  Ali Thabet}\\
	Visual Computing Center\\
	KAUST\\
	Thuwal, SA\\ 
	\texttt{ali.thabet@kaust.edu.sa}\\
	\And
	\href{https://orcid.org/0000-0002-4227-8508}{\includegraphics[scale=0.06]{orcid.pdf}\hspace{1mm}  Wolfgang Heidrich}\\
	Visual Computing Center\\
	KAUST\\
	Thuwal, SA\\ 
	\texttt{wolfgang.heidrich@kaust.edu.sa}\\}
	
\renewcommand{\shorttitle}{Snapshot HDR Video Construction Using a Coded Mask}

\hypersetup{
pdftitle={A template for the arxiv style},
pdfsubject={q-bio.NC, q-bio.QM},
pdfauthor={David S.~Hippocampus, Elias D.~Striatum},
pdfkeywords={First keyword, Second keyword, More},
}

\begin{document}
\maketitle

\begin{abstract}
This paper study the reconstruction of High Dynamic Range (HDR) video from snapshot-coded LDR video. Constructing an HDR video requires restoring the HDR values for each frame and maintaining the consistency between successive frames. HDR image acquisition from single image capture, also known as snapshot HDR imaging, can be achieved in several ways.  For example, the reconfigurable snapshot HDR camera is realized by introducing an optical element into the optical stack of the camera; by placing a coded mask at a small standoff distance in front of the sensor. High-quality HDR image can be recovered from the captured coded image using deep learning methods. This study utilizes 3D-CNNs to perform a joint demosaicking, denoising, and HDR video reconstruction from coded LDR video. We enforce more temporally consistent HDR video reconstruction by introducing a temporal loss function that considers the short-term and long-term consistency. The obtained results are promising and could lead to affordable HDR video capture using conventional cameras. 

\end{abstract}

% keywords can be removed

\section{Introduction}
%\lipsum[2]
%\lipsum[3]
The human visual system can sense up to 20 F-stops of luminance contrast with minimal eye adaption~\cite{artusi2017advanced}. Modern image sensor technology is incapable of matching this performance and reproducing the full dynamic range of natural scenes within a single exposure. The challenge of single shot (or snapshot) High Dynamic Range (HDR) imaging arises from the tremendous gap between the huge intensity range in natural scenes and the very limited bit depths that modern camera sensors can offer. 

Modern films %have often been shot with cameras having 
are often shot using cameras with a higher dynamic range%. It 
, which mainly require both HDR shooting and rendering in addition to special effects, particularly seamless mixing of natural and synthetic footage. HDR video is also required in all applications that require high accuracy in capturing temporal aspects of the changes in a scene. HDR video
capture is essential for specific industrial tracking processes, such as melting, machine vision such as autonomous driving, and monitoring systems.

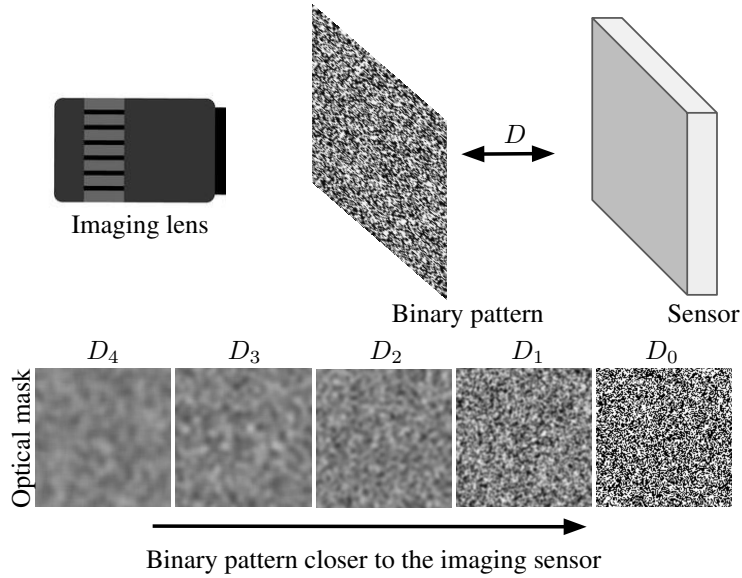
\begin{figure}[!h]
	\centering
	%\vspace{-2 mm}
	\begin{tikzpicture}
	\node[anchor=south west,inner sep=0] (image) at (0,0) 
	{\begin{overpic}[width=.6\columnwidth]{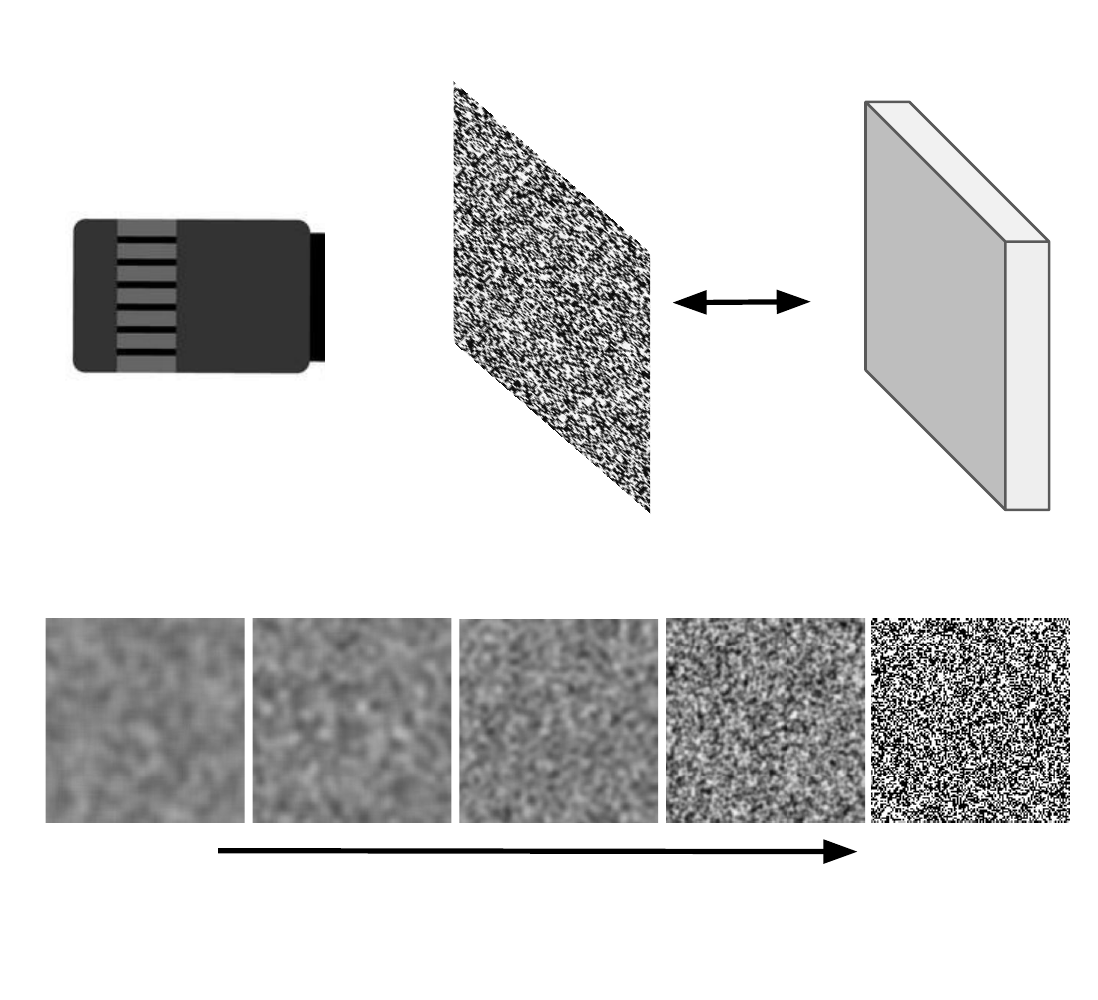}
		\put(11,36){{$D_4$}}
		\put(30,36){{$D_3$}}
		\put(49,36){{$D_2$}}
		\put(68,36){{$D_1$}}
		\put(86,36){{$D_0$}}
		
		\put(67,65){{$D$}}
		\put(89,41){Sensor}
		\put(52,41){Binary pattern}
		\put(9,53){Imaging lens}
		\put(1,14.5){\rotatebox{90}{\, {Optical mask}}}	
		\put(19,8){Binary pattern closer to the imaging sensor}
		
		\end{overpic}};
	%\begin{scope}[x={(image.south east)},y={(image.north west)}]
	%\draw[color=green, thick] (0.001,0.53) rectangle ++(0.055,0.17);
	%\draw[color=green, thick] (0.23,0.68) rectangle ++(0.055,0.17);
	%\draw[color=yellow, thick] (0.135,0.095) rectangle ++(0.055,0.17);
	%\draw[color=yellow, thick] (0.06,0.015) rectangle ++(0.055,0.17);
	%\end{scope}
	\end{tikzpicture}
	%\vspace{-10 mm}
	\caption[Effect of binary pattern distance from the sensor plane on the optical mask.]{Illustration of the effect of the binary pattern distance from the sensor plane on the resulting optical mask. When the binary pattern is exactly on the sensor plane, we obtain the exact binary pattern mask. As the binary pattern is moved away from the sensor by various distances $D$, the mask becomes blurred versions of the binary pattern. The blurriness depends on the distance.}
	\label{fig:MaskSimulation}
	%\vspace{-5 mm}
\end{figure}
In \cite{alghamdi2019,alghamdi2021Transfer} we proposed a computational imaging solution to single shot HDR imaging by minimal modifications of the camera to implement per-pixel exposure, as well as a deep learning algorithm based on the inception network to reconstruct HDR images. Specifically we explored a variant of the spatially modulated HDR camera design that does not require a custom sensor, and can be incorporated into any existing camera, be it a smartphone, a machine vision camera, or a digital SLR. We envision in particular a scenario where the camera can be reconfigured on the fly into an HDR mode with the introduction of an optical element into the optical stack of the camera. To realize these desired properties, we propose a mask that is not attached directly to the surface of the image sensor as in the case of Assorted Pixels~\cite{nayar2000high,nayar2003adaptive}, but is instead placed at a small standoff distance in front of the sensor. To support dynamic hardware reconfiguration, we explored rapid calibration of the mask, as well as snapshot HDR image reconstruction.  We proposed (1)an easy-to-implement modulation method that requires minimum hardware modification and a simple self-calibration technique; (2) a new HDR reconstruction algorithm built upon inception network that decodes decent HDR images from the raw Bayer data. We demonstrated both in simulation and by a prototype that the combination of hardware encoding and software. In \cite{alghamdi2019,alghamdi2021Transfer}, our primary focus was %on 
capturing and reconstructing a single HDR image from a single coded LDR %I
image, and we worked mainly on the spatial domain of the image.

Our main focus in this paper is to work with coded LDR videos, where we expect to benefit greatly from working jointly on both the spatial and temporal domains of video images. In this article, we will concentrate on  constructing HDR video from coded LDR images obtained using the reconfigurable snapshot HDR camera. Reconstructing HDR video from coded LDR video requires restoration of the HDR values for each frame and maintenance of the consistency between successive frames. Spatio-temporal coherence between, and among, the frames must be exploited appropriately for accurate prediction of HDR values. Although 2D-CNNs are powerful for modeling images, 3D-CNNs are more appropriate for spatio-temporal feature extraction as they can maintain the temporal information. For this reason, the reconstruction would fail if we directly used our proposed networks in \cite{alghamdi2019, alghamdi2021Transfer} to video frames separately, because they lack a mechanism to preserve temporal coherence.

\section {Related Work}
Basically there are two main methods for acquiring HDR images%exist
: HDR sensors, and multiple LDR exposures captured with standard sensors. Some HDR cameras %were
have been presented to the scientific society, %and 
but are still not accessible for market consumers, e.g.,~\cite{nayar2003adaptive,tocci2011versatile,chalmers2011hdr}. There %are 
are a few alternative commercial HDR sensors, such as the Red Epic camera~\cite{RedCompany}, Thomson Viper~\cite{GrassValley}, Arri Alexa~\cite{ArriAlexa}, Sony PXW-Z90~\cite{Sony}, and Phantom HD \cite{VisionResearch}, %which 
though these devices still have a limited dynamic range under 16 f-stops and %way too 
are extremely expensive.   
To overcome this limitation, many computational imaging techniques have been developed via co-designing the sensor architecture and post-processing algorithms for HDR image acquisition~\cite{reinhard2010high}. These methods can be categorized into three distinct approaches. The most common way is to capture a sequence of low dynamic range (LDR) images with different exposures and fuse them into an HDR image~\cite{debevec1997recovering,Picard1995onbeing}. Modern cameras and mobile devices can easily afford successive image capture, making this method capable of producing decent HDR images for static scenes. However, when either the scene is dynamic or the camera shakes during capture, the resulting images can suffer from ghosting artifacts. The second approach is to utilize multiple sensors to simultaneously capture differently exposed LDR images by, for example, splitting the light to multiple sensors with a beam-splitter~\cite{mcguire2007optical,tocci2011versatile,kronander2013unified}. This sophisticated approach is expensive and needs additional rigorous calibration. The third approach is to capture a single LDR image with a per-pixel or per-scanline coded exposure. Reconstruction algorithms are applied later to create HDR images~\cite{nayar2000high,nayar2003adaptive,serrano2016convolutional}. This type of computational camera can be achieved by using a per-pixel coded exposures in the sensor architecture~\cite{kensei2014image} or by mounting an optical mask onto an off-the-shelf camera sensor.

In \cite{alghamdi2019} for easy implementation of a grayscale mask, we choose to place a random binary optical mask at a short distance (typically 1-2 mm) in front of the sensor. Note that we did not optimize the distance, but simply mounted our mask on the cover glass that is usually present in front of the sensor. Light propagation from the mask to the sensor results in a blurred version of the binary mask. The actual statistics depends on both the mask and propagation distance. Figure~\ref{fig:MaskSimulation} illustrates the effect of distance on the resulting optical mask. For HDR reconstruction we introduced an algorithm built upon an inception network that decodes reliable HDR images from the raw noisy coded Bayer data. We demonstrate both in simulation and using a prototype that the combination of hardware encoding and software decoding leads to a simple, yet efficient, HDR image acquisition system. 

In \cite{alghamdi2021Transfer} we present a transfer learning framework for solving the HDR reconstruction part.  Our motivation comes from the fact that available HDR image datasets are small compared to the typical requirement for training deep neural networks. In ~\cite{alghamdi2019} we solved this issue by pre-training on a large simulated HDR dataset. This pre-training is expensive in both in memory and time; experimenting with different network structures will need weeks of pre-training. In tIn \cite{alghamdi2021Transfer} we incorporate architectures pre-trained on a different large scale task, and transfer them to our HDR reconstruction. This new approach reduces our processing time substantially. Specifically, we  propose an encoder-decoder framework, that learns an initial estimation of the HDR image, as well as useful image features. We then refine our estimate through residual learning \cite{ronneberger2015u}. Our final network can be trained end-to-end. For the encoder, we use a VGG16 \cite{simonyan2014very} network pre-trained on ImageNet. With few epochs of training on a small dataset the network learned to reconstruct high quality results.

3D-CNNs have successfully been applied to high-level vision tasks for videos, such as action recognition and event classification~\cite{ji20123d, tran2015learning}. The spatio-temporal feature extraction capability of 3D-CNNs was demonstrated in~\cite{ji20123d, tran2015learning}. In~\cite{tran2015learning}, the authors argued that 3D-CNNs provide an adequate video descriptor, and a homogeneous architecture with small 3×3×3 convolution kernels in all layers is among the best-performing architecture for 3D-CNNs. Moreover, the capabilities of 3 D-CNN in video enhancement, inpainting and super-resolution have been proven~\cite{lv2018mbllen, Kappeler2016VideoSuper, Wang_2017_ICCV, Wang_Huang_Han_Wang_2019}. This article will use a 3D CNN to globally perform a joint demosaicking, denoising, and HDR video reconstruction coded LDR video. As far as the author knows, there is no published work on the construction of HDR video from coded LDR images that utilizes temporal information in the reconstruction process.

\section {Methods}
\subsection {Imaging Model}

In our HDR system, we propose placing an optical mask into the optical path in close proximity to the image sensor. The propagation of light from the mask to the sensor leads to a grayscale modulation pattern on the captured image. In a color camera, a Bayer Color Filter Array (CFA) samples the radiance into three color channels. The camera sensor then converts the photons impinging on the image plane over a specific exposure time into electrons, and quantizes the voltage values into digital numbers (DNs). Basically, the process of capturing coded LDR video can be mathematically expressed as %follow:
follows:

\begin{equation}
\Vect{y_k} = g \left( f \left( \Mat{B} \Phi \Vect{x_k} \Delta t \right) \right),        k=1,2,3,..
\label{eq5:video_model}
\end{equation}  

where, $\Vect{y_k} \in \mathbb{R}^M$ is the captured raw LDR image, $\Vect{x_k} \in \mathbb{R}^M$ is the radiance in the scene, $k$ is the frame/time-instance index, $\Phi \in \mathbb{R}^{M \times M}$ is the spatially varying modulation mask, $\Mat{B} \in \mathbb{R}^{M \times M}$ is a diagonal matrix that represents the Bayer filter, and $\Delta t$ is the exposure time. $f$ is a nonlinear function that includes the camera response and quantization. $g$ is a noise function that accounts for different types of noise in modern imaging sensors, including photon shot noise, dark current, fixed pattern noise, quantization noise and other nonlinearities~\cite{konnik2014high}.

\subsection {HDR Reconstruction Network}

\begin{figure*}[h!]
	\centering	
	%\includesvg [width=.99\textwidth]{images/Encoder_Decoder_enhancer.svg}
	%\vspace{ 3mm}
	\includegraphics[width=1\linewidth]{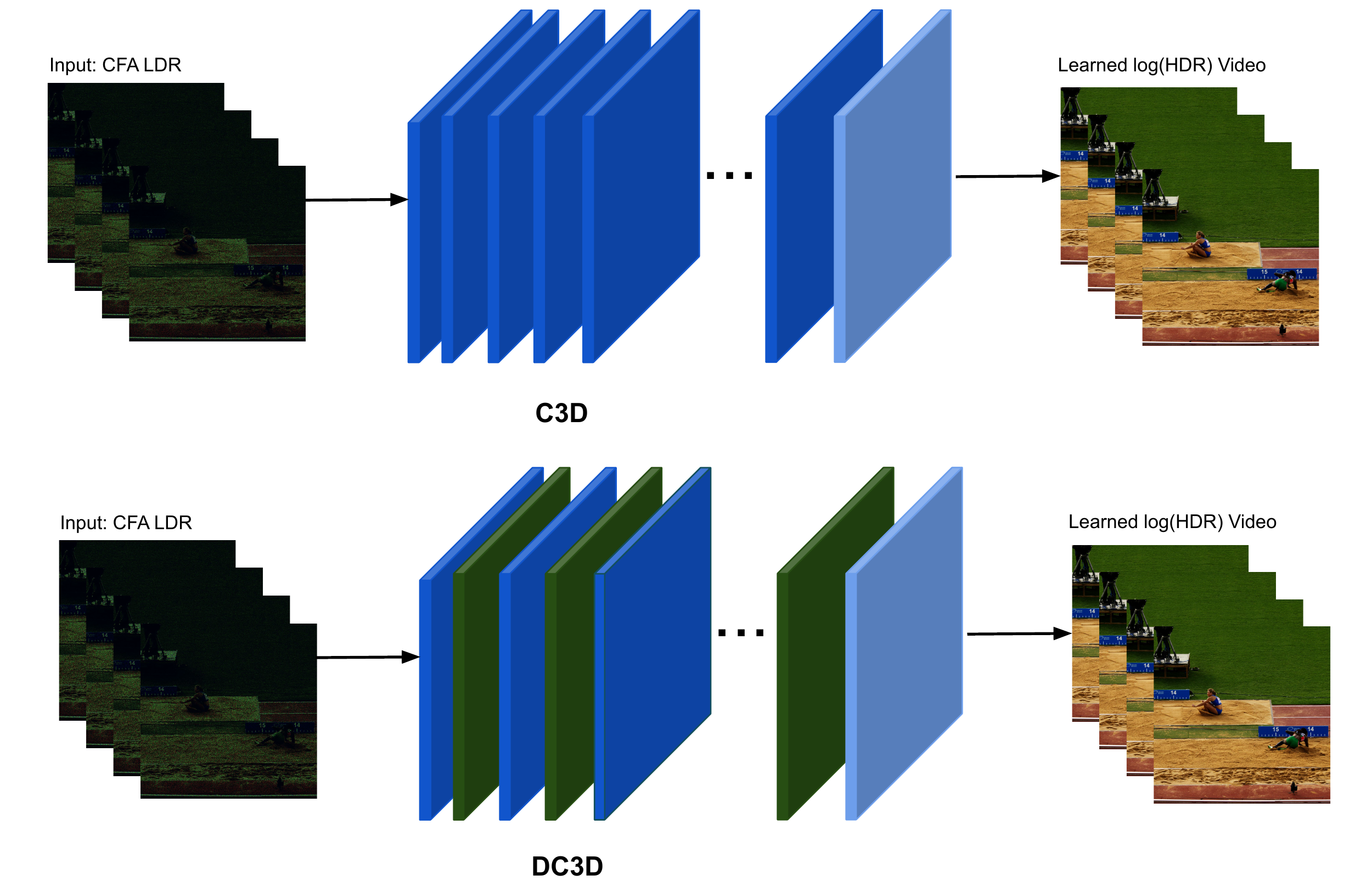}
	\\
	
	\vspace{- 3 mm}
	%\hspace{-20mm}	
	\includegraphics[width=1\linewidth]{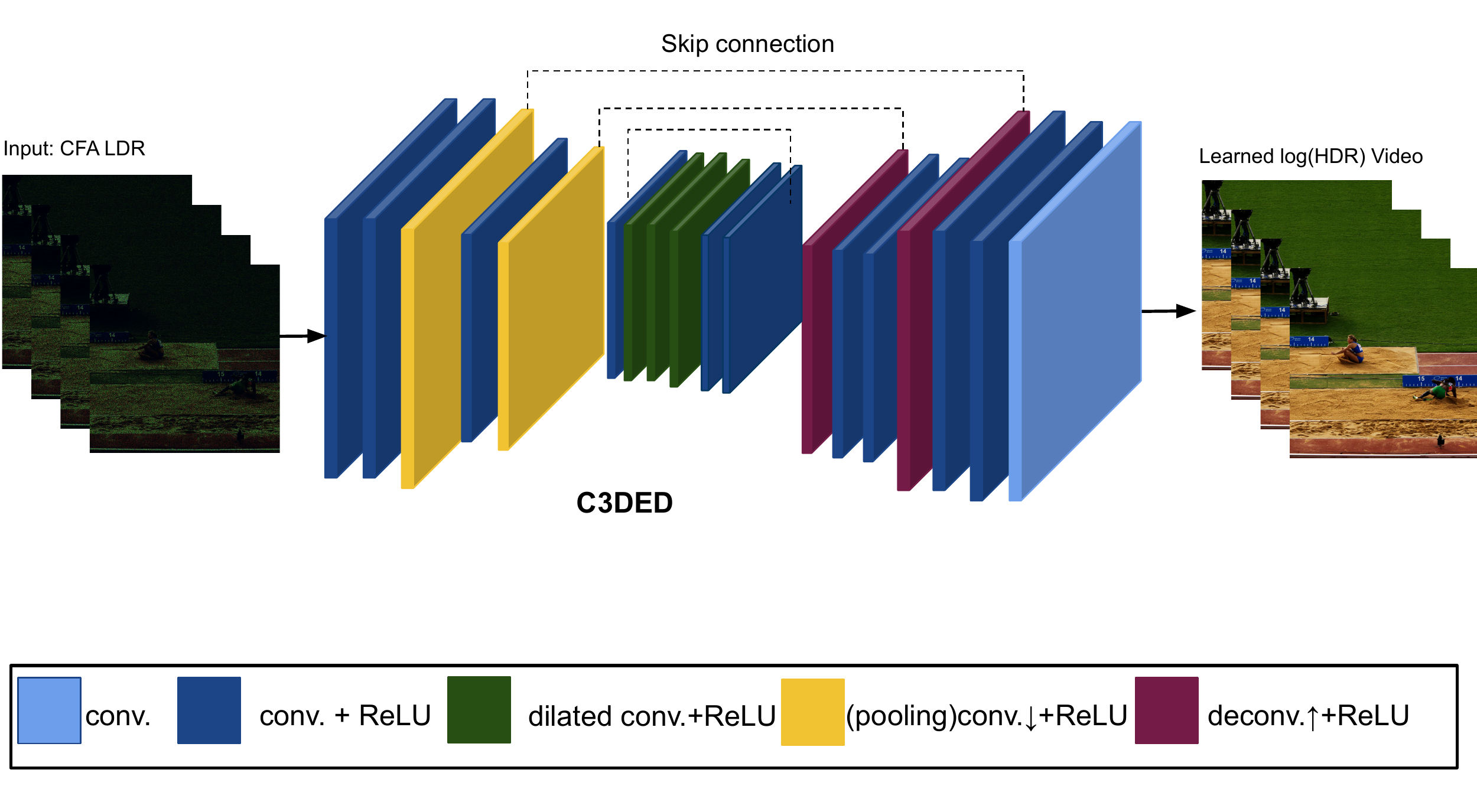}
	%	\def\svgwidth{0.9\textwidth}\scriptsize
	%{\fontsize{9}{9}\input{../VMV/images/HDR_Reconstruction_Inception_Netwo%rk4.pdf_tex}}\\
	%\def\svgwidth{.99\textwidth}
	%\input{images/HDR_Reconstruction_Inception_Network4.pdf_tex}     
	\vspace{-3 mm}
	\caption[ 3D convolutional networks for Video HDR Reconstruction]{Examined 3D convolutional networks for Video HDR Reconstruction. All networks take as input the captured raw LDR frames $y$ and output an estimation for the $log(HDR)$ video frames $\hat{x}$.}
	\label{ch6fig:HDR_Reconstruction_Nets_vid}
	\vspace{-3mm}
	\end {figure*}

We experimented with three different 3D convolutional networks, as illustrated in Figure~\ref{ch6fig:HDR_Reconstruction_Nets_vid}. All models have the same input, which is the coded CFA LDR video frames, $y \in \mathbb{R}^{f\times h \times w \times c}$, where $f$ represents the number of consecutive frames, $h$ and $w$ indicate height and width, $c=3$ for $RGB$ channels, and output the estimated HDR video frames $x \in \mathbb{R}^{f\times h \times w \times c}$.

\textbf{C3D} is a simple network that consists of ten convolutional layers without any pooling or dilation. Each convolutional layer is followed by a ReLU non-linearity activation, except the last layer. Paddings are used to force the input and output to have the same size. The detailed configuration of the C3D network is illustrated in Figure~\ref{ch6fig:HDR_Reconstruction_Nets_vid} and listed in Table~\ref{tb:c3d}.

The second model is a dilated convolutional 3D network~\textbf{DC3D}; this model utilizes dilated convolutions, which allows the receptive field to expand exponentially without losing resolution~\cite{yu2015multi}. The  DC3D also consists of ten convolutional layers, but the second, fourth, seventh, and ninth layers are dilated convolutional layers. Each convolutional layer is followed by a ReLU non-linearity activation, except the last %one
layer. Paddings are used to %enforce
force the input and output to have the same size. The detailed configuration of the DC3D network is illustrated in Figure~\ref{ch6fig:HDR_Reconstruction_Nets_vid} and listed in Table~\ref{tb:C3D_dil}.

Our \textbf{C3DED} network inspired by ~\cite{Wang_Huang_Han_Wang_2019} follows an encoder-decoder structure and consists of 18 layers. Given the coded CFA video, the network first consists of six convolutional layers. The third and fifth layers are strided convolutional layers to encode feature-maps to a latent space, capturing its temporal-spatial structure. Then, three dilated convolutional layers %with rate two 
with a rate of two are employed to capture the spatial-temporal information in a larger perception field. Finally, HDR video is achieved by seven convolutional layers and two fractionally-strided convolutional layers. Their order is illustrated in Table~\ref{tb:C3DED}. To ensure that every pixel contributes, we use 3x3 convolution kernels with a stride of 2, instead of max-pooling and upsampling layers, to compute the feature maps. Recognizing that non-successive frames may have relations with the current frame and avoid information loss beyond frames, we restrict stride and dilation to impact only within frames, rather than across frames; a similar approach was used in~\cite{Wang_Huang_Han_Wang_2019}. As a result, the feature map of each layer has a constant frame number f.  Skip-connections, as in U-Net~\cite{ronneberger2015u}~, are also employed to facilitate feature compounding between the encoder and decoder. Moreover, %all 
all of the convolutional layers are followed by a ReLU non-linearity activation, except the last. Paddings are used to force the input and output to have the same size. The detailed configuration of our 3D completion network is illustrated in Figure~\ref{ch6fig:HDR_Reconstruction_Nets_vid} and listed in Table~\ref{tb:C3DED}.  
\begin{table}
	
	\caption[C3D network architecture.]{C3D network architecture. All convolutional layers are followed by ReLU, except for the last layer. K: kernels, S: strides, Ch: output channels, D: dilation.}
	
	\centering
	\begin{tabular}{|l|lllll||l|lllll|} 
		
		\hline
		L  & Type  & K & S & Ch & D & L & Type  & K & S & Ch & D  \\ 
		\hline
		1    & conv. & 3   & 1   & 64   & -        & 6   & conv. & 3   & 1   & 64   & -         \\
		2    & conv. & 3   & 1   & 64   & -        & 7   & conv. & 3   & 1   & 64   & -         \\
		3    & conv. & 3   & 1   & 64   & -        & 8   & conv  & 3   & 1   & 64   & -         \\
		4    & conv. & 3   & 1   & 64   & -        & 9   & conv. & 3   & 1   & 64   & -         \\
		5    & conv. & 3   & 1   & 64   & -        & 10  & conv. & 3   & 1   & 3   & -         \\
		\hline
	\end{tabular}
	\vspace{-2mm}
	\label{tb:c3d}
\end{table}

\begin{table}
	\caption[DC3D Network architecture]{DC3D Network architecture. All convolutional layers are followed by ReLU, except for the last layer. K: kernels, S: strides, Ch: output channels, D: dilation.}
	\centering
	\begin{tabular}{|l|lllll||l|lllll|} 
		\hline
		L  & Type  & K & S & Ch & D & L & Type  & K & S & Ch & D  \\ 
		\hline
		1    & conv. & 3   & 1   & 64   & -        & 6   & conv. & 3   & 1   & 64   & -         \\
		2    & dilated conv. & 3   & 1   & 64   & 2        & 7   & dilated conv. & 3   & 1   & 64   & 2         \\
		3    & conv. & 3   & 1   & 64   & -        & 8   & conv  & 3   & 1   & 64   & -         \\
		4    & dilated conv. & 3   & 1   & 64   & 2        & 9   & dilated conv. & 3   & 1   & 64   & 2         \\
		5    & conv. & 3   & 1   & 64   & -        & 10  & conv. & 3   & 1   & 3   & -         \\
		\hline
	\end{tabular}
	\label{tb:C3D_dil}
\end{table}
\begin{table}
	\caption[C3DED network architecture]{C3DED network architecture. All convolutional layers are followed by ReLU, except for the last layer. K: kernels, S: strides, Ch: output channels, D: dilation.}
	\centering
	\refstepcounter{table}
	\label{tb:C3DED}
	\begin{tabular}{|l|lllll||l|lllll|} 
		\hline
		L                  & Type               & K & S     & Ch & D             & L                  & Type               & K & S     & Ch & D              \\ 
		\hline
		1                    & conv.              & 5   & 1       & 16   & -                    & 10                   & conv.              & 3   & 1       & 128  & -                     \\
		2                    & conv.              & 5   & 1       & 16   & -                    & 11                   & conv.              & 3   & 1       & 64   & -                     \\
		3                    & conv.$\downarrow$  & 3   & (1,2,2) & 32   & -                    & 12                   & deconv.$\uparrow$   & 5   & (1,2,2) & 32   & -                     \\
		4                    & conv.              & 3   & 1       & 64   & -                    & 13                   & conv.              & 3   & 1       & 64   & -                     \\
		5                    & conv.$\downarrow$  & 3   & (1,2,2) & 128  & -                    & 14                   & conv.              & 3   & 1       & 32   & -                     \\
		6                    & conv.              & 3   & 1       & 128  & -                    & 15                   & deconv.$\uparrow$  & 3   & (1,2,2) & 16   & -                     \\
		7                    & dilated conv.      & 3   & (1,2,2) & 128  & 2                    & 16                   & conv.              & 3   & 1       & 32   & -                     \\
		8                    & dilated conv.      & 3   & (1,2,2) & 128  & 2                    & 17                   & conv.              & 3   & 1       & 16   & -                     \\
		9                    & dilated conv.      & 3   & (1,2,2) & 128  & 2                    & 18                   & conv.              & 3   & 1       & 3    & -                     \\ 
		\hline
		\multicolumn{1}{l}{} &                    &     &         &      & \multicolumn{1}{l}{} & \multicolumn{1}{l}{} &                    &     &         &      & \multicolumn{1}{l}{}  \\
		\multicolumn{1}{l}{} &                    &     &         &      & \multicolumn{1}{l}{} & \multicolumn{1}{l}{} &                    &     &         &      & \multicolumn{1}{l}{} 
	\end{tabular}
\end{table}

\subsection{Loss function}

Our goal is to reconstruct reliable HDR frames that look like ground-truth frames, and we want the reconstructed frames to be temporally consistent. Therefore, we propose to train our models with (1)  content loss between the ground-truth HDR frames $x$ and output $\hat{x}$ learned by the network, and ~(2) short-term and long-term temporal losses between output frames $\hat{x}$.

\noindent\textbf{Content loss:} 
This loss consists of two parts: (1) a pixel-wise image domain mean L1 error (MAE) loss and (2) a perceptual VGG  loss, and %both 
both are computed for every frame $f$.

\begin{equation}
\label{eq:ch6_content_loss}
\begin{gathered}
L_{Content} \left( x,\hat{x} \right) =\lambda_{1} L_1 \left( x, \hat{x}\right) + \lambda_{2}  L_{VGG} \left( x, \hat{x}\right),
\end{gathered}
%\vspace{-2mm}
\end{equation}

\noindent Where the perceptual loss $L_{VGG}$ %defined
is defined as follows, given a network $V$ (e.g. VGG), 
% we compute the distance between reference $x$ and learned (or test) image \change{$\hat{x}$} as follows. 
%\change{
we compute the loss between each learned image $\hat{x}$ and its corresponding ground truth $x$, as follows: %}
First, we utilize $V$ as a feature extractor, by selecting the output of $N$ layers. Finally, we define our new loss as in Eq.\eqref{eq5:perceptual_loss}, where %~\change{
$x$ is the logarithm of the ground truth HDR image, $\hat{x}$ is the output of the last layer of our HDR reconstruction network, and ~$x^{i}$ and $\hat{x}^{i}$ $\in \mathbb R^{H^i\times W^i\times C^i}$ are outputs from the $i$-th layer of network $V$. 
\begin{equation}
\label{eq5:perceptual_loss}
L_{VGG}(x,\hat{x}) = \sum_{i=1}^{i=N} \frac{1}{H^i\times W^i\times C^i} \omega_{i} \left\Vert x^i- \hat{x}^i \right\Vert_1,
\end{equation}
where $\omega_{i}$ is used for weighting the contribution of features extracted from layer $i$ to the loss function. Our final loss function can be defined as 

\noindent\textbf{Temporal loss:} 
Since we use supervised training to train our models, we can force the temporally coherent pixels in the ground-truth~$x$ to be temporally coherent in the recovered output~$\hat{x}$. Our temporal loss consists of two parts: (1) short-term memory loss, where the motivation is to enforce consecutive frames to be temporally coherent, and (2) long-term memory loss to enforce long-term temporal coherence, the simplest way is to enforce all frames to be temporally consistent with the first frame. 

\noindent\textbf{Short memory loss} can be defined using the following equation:

\begin{equation}
\label{eq:ch6_shortMem_loss}
\begin{gathered}
L_{SM} \left( x,\hat{x} \right) = \frac{1}{f-1} \sum_{t=2}^{f} \lVert  \left( w_{t\rightarrow t-1} \odot \left( \hat{x}_t - \hat{x}_{t-1} \right)  \right) \rVert_{1},
\end{gathered}
%\vspace{-2mm}
\end{equation}

\noindent the~$\odot$ represents the pixel-wise product, and $w_{t}$ is the visibility weight matrix calculated from the warping error between the ground-truth HDR frame $x_{t}$ and $x_{t-1}$. The visibility weighting matrix between two frames is computed as follows:

\begin{equation}
\label{eq:ch6_w_visibility}
\begin{gathered}
w_{v\rightarrow u}^{ij}= exp\left(-\tau\left( x_{u}^{ij} - x_{v}^{ij}\right)\right) ,
\end{gathered}
%\vspace{-2mm}
\end{equation}

\noindent where $ij$ %indicate 
indicates the pixel's spatial location, $u$ and $v$ indicate the temporal index of the frame. $\tau>1$, we used $\tau=100$ in our experiments,  $w_{v\rightarrow u}^{ij}$ %are high approaching 
approaches 1 when the ground-truth pixels $x_{u}^{ij}$ and $x_{v}^{ij}$ are similar. On the contrary, when they are different, the value of $w_{v\rightarrow u}^{ij}$ %approach 
approaches 0.

\noindent Similarly, \textbf{Long-term memory loss} can be defined using the following equation: 

\begin{equation}
\label{eq:ch6_LongMem_loss}
\begin{gathered}
L_{LM} \left( x,\hat{x} \right) = \frac{1}{f-1} \sum_{t=2}^{f} \lVert  \left( w_{t\rightarrow 1} \odot \left( \hat{x}_t - \hat{x}_{1} \right)  \right) \rVert_{1},
\end{gathered}
%\vspace{-2mm}
\end{equation}

\noindent Now we can define %our 
the temporal loss as %follow
follows:

\begin{equation}
\label{eq:ch6_Tem_loss}
\begin{gathered}
L_{T} \left( x,\hat{x} \right) = \lambda_{3} L_{SM} \left( x,\hat{x} \right)+\lambda_{4} L_{LM} \left( x,\hat{x} \right)  ,
\end{gathered}
%\vspace{-2mm}
\end{equation}

\noindent and from this, our loss function can be defined as %follow
follows:

\begin{equation}
\label{eq:ch6_loss}
\begin{gathered}
L_{VHDR} \left( x,\hat{x} \right) = \lambda_{1} L_1 \left( x, \hat{x}\right) + \lambda_{2}  L_{VGG} \left( x, \hat{x}\right)+\lambda_{3} L_{SM} \left( x,\hat{x} \right)+\lambda_{4} L_{LM} \left( x,\hat{x} \right),
\end{gathered}
%\vspace{-2mm}
\end{equation}

\section{Training}
To train our networks, we combined HDR video from three publicly available HDR video datasets: DML-HDR~\cite{banitalebi2014compression}, LiU HDRv \cite{kronander2013unified}, and the Zurich Athletics 2014 dataset \cite{EBU_2014}. We obtained $63$ shots, with a total of $59K$ HDR video frames. We used $45$ shots with around $46K$ HDR frames to synthesize the training and validation dataset. We used the remaining $18$ shots with a total of $13K$ frames to generate the testing data. We synthesized the coded CFA LDR video frames $y$ in a similar manner as the still data \cite{alghamdi2019, alghamdi2021Transfer}. However, we fixed the exposure time to  $.033 ms$ to simulate $30 fps$ video capture. As in the still case, we simulate both the ideal sharp mask statistics and the more realistic low-frequency masks resulting from a non-zero spacing between the mask and the image plane. Both are shown in Figure~\ref{fig:MaskTypeEffect}. As a result, 50 percent of the simulated masks are uniform masks, where each mask pixel has a random value between zero and one. The other 50 percent are LF Gaussian masks, which simulate the finite distance between the binary pattern and the sensor in our current prototype, as explained in Section~\ref{subsec3:ExperminatalResults}. We simulated the fabricated binary pattern as a random Bernoulli matrix of zeros and ones. More information about the mask simulation and synthesized LDR generation can be found in section~\ref{ch3:Training}.

We used the $45$ HDR shots to generate $3K$ coded LDR samples for the training set, each of which %consists
consist of $f=8$ frames and $h=w=512$. We selected a random training HDR video for each sample, randomly %pick 
picked a starting frame, then chose a random crop location. We fed the selected frames into the simulation module. We trained the HDR reconstruction networks using PyTorch~\cite{paszke2017automatic} with two NVIDIA V100 GPUs. For learning, we used the ADAM optimizer~\cite{kingma2014adam}. We set the learning rate to $10^{-4}$. All models %trained for 
were trained over $600$ epochs, with a mini-batch size of 4. We set $\lambda_1=1$, then conducted experiments with $\lambda_2=[0.01,0.1]$ , and $\lambda_3=\lambda_4=[0.001, 0.01, 0.1, 1]$. For each model, we trained ten networks; two utilize $L_{Content}$ loss, and eight use $L_{VHDR}$ loss.

Here, we %want to 
clarify that the purpose of the experiments presented in this paper is to estimate the performance of 3D models and the loss functions introduced in the previous section. $3K$ samples are by no means enough for training image reconstruction CNN. We %experiment 
experimented with training our models %with 
on $30K$ versus $3K$ samples for 100 epochs, and the results were numerically and visually similar. A final model should be trained on a larger dataset.

\section{Results}

\begin{table}[]
	\centering
		 
	\caption[Evaluation of 3D models on HDR video reconstruction.]{Evaluation of three 3D convolutional models. For each model, we reported two loss functions. We tested 200 coded LDR clips simulated %form
	from the test shots. We computed the average HDR-VDP2 $Q_{Score}$\cite{mantiuk2011hdr}; larger values are better (up to 100). Also we %compute 
	computed the mean absolute error (MAE), mean squared error (MSE), temporal loss $L_T$\eqref{eq:ch6_Tem_loss}, SSIM \cite{wang2004image}, and perceptual loss. $L_{VGG}$\eqref{eq5:perceptual_loss}.
		} 
	\label{ch6:sim_res}
	\begin{tabular}{l|l|l|l|l|l|l|}
		\cline{2-7}
		& \multicolumn{2}{c|}{\textbf{C3D}} & \multicolumn{2}{c|}{\textbf{DC3D}} & \multicolumn{2}{c|}{\textbf{C3DED}} \\ \cline{2-7} 
		& $L_{Content}$    & $L_{VHDR}$     & $L_{Content}$    &$L_{VHDR}$             & $L_{Content}$             & $L_{VHDR}$             \\ \hline
		\multicolumn{1}{|l|}{Qscore} & 53.34 & 53.58  & 53.63 & 55.03          & 54.93          & \textbf{55.07} \\
		\multicolumn{1}{|l|}{MAE}     & 0.087 & 0.074  & 0.079 & \textbf{0.064} & 0.077          & 0.067          \\
		\multicolumn{1}{|l|}{MSE}      & 0.016           & 0.011           & 0.011       & \textbf{0.010}       & 0.012            & \textbf{0.010}   \\
		\multicolumn{1}{|l|}{$L_T$\eqref{eq:ch6_Tem_loss}} & 0.145           & 0.135           & 0.131       & 0.128                & \textbf{0.123}   & \textbf{0.123}   \\
		\multicolumn{1}{|l|}{SSIM}   & 0.784 & 0.798 & 0.804 & 0.832          & \textbf{0.847} & 0.837          \\
		\multicolumn{1}{|l|}{$L_{VGG}$\eqref{eq5:perceptual_loss}}    & 0.102 & 0.100  & 0.104 & 0.094          & 0.096          & \textbf{0.093} \\ \hline
	\end{tabular}
\end{table}
 
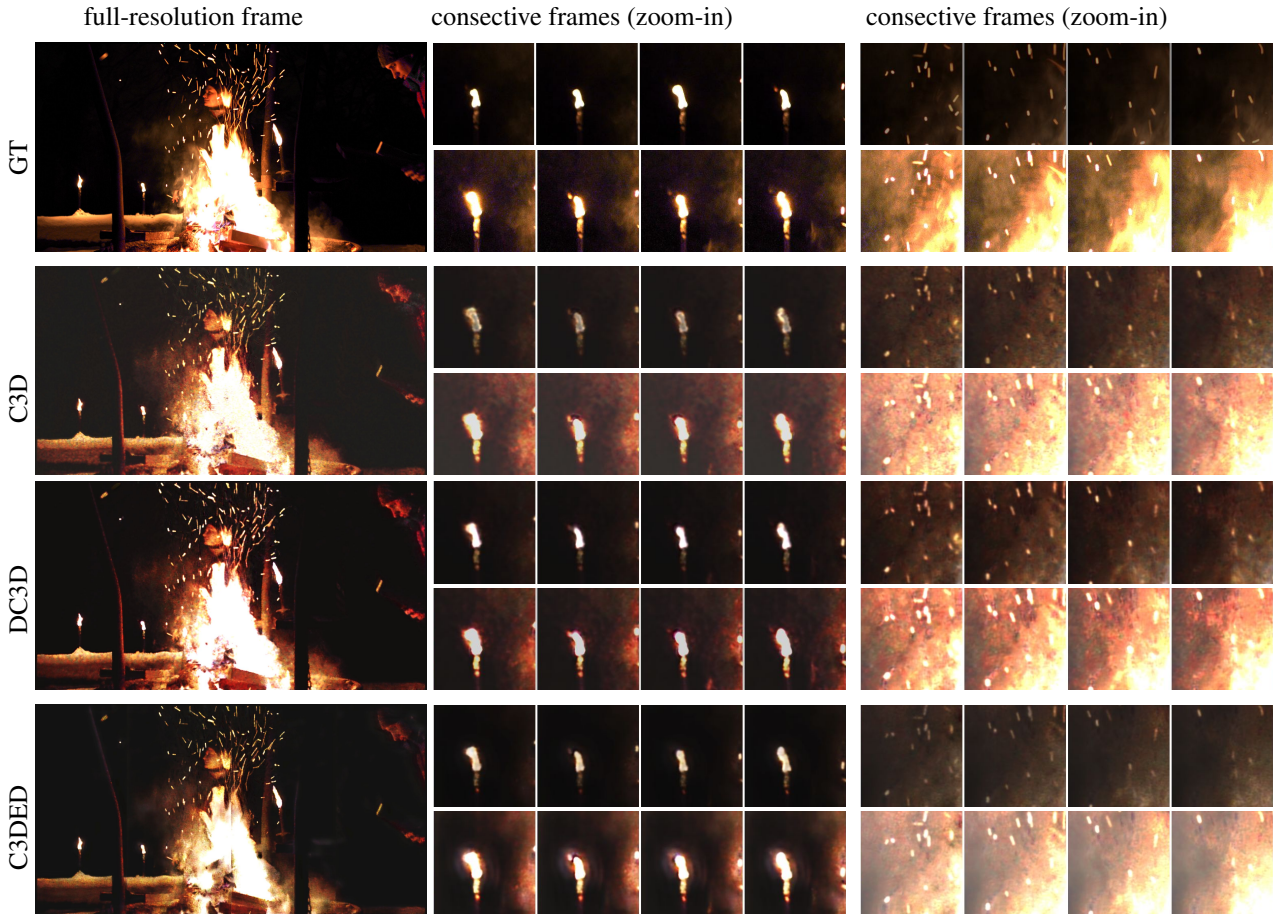
\begin{figure}[hbt!]
	\centering
	\vspace{3 mm}
	\begin{tikzpicture}
	\node[anchor=south west,inner sep=0] (image) at (0,0) 
	{\begin{overpic}[width=1\columnwidth]{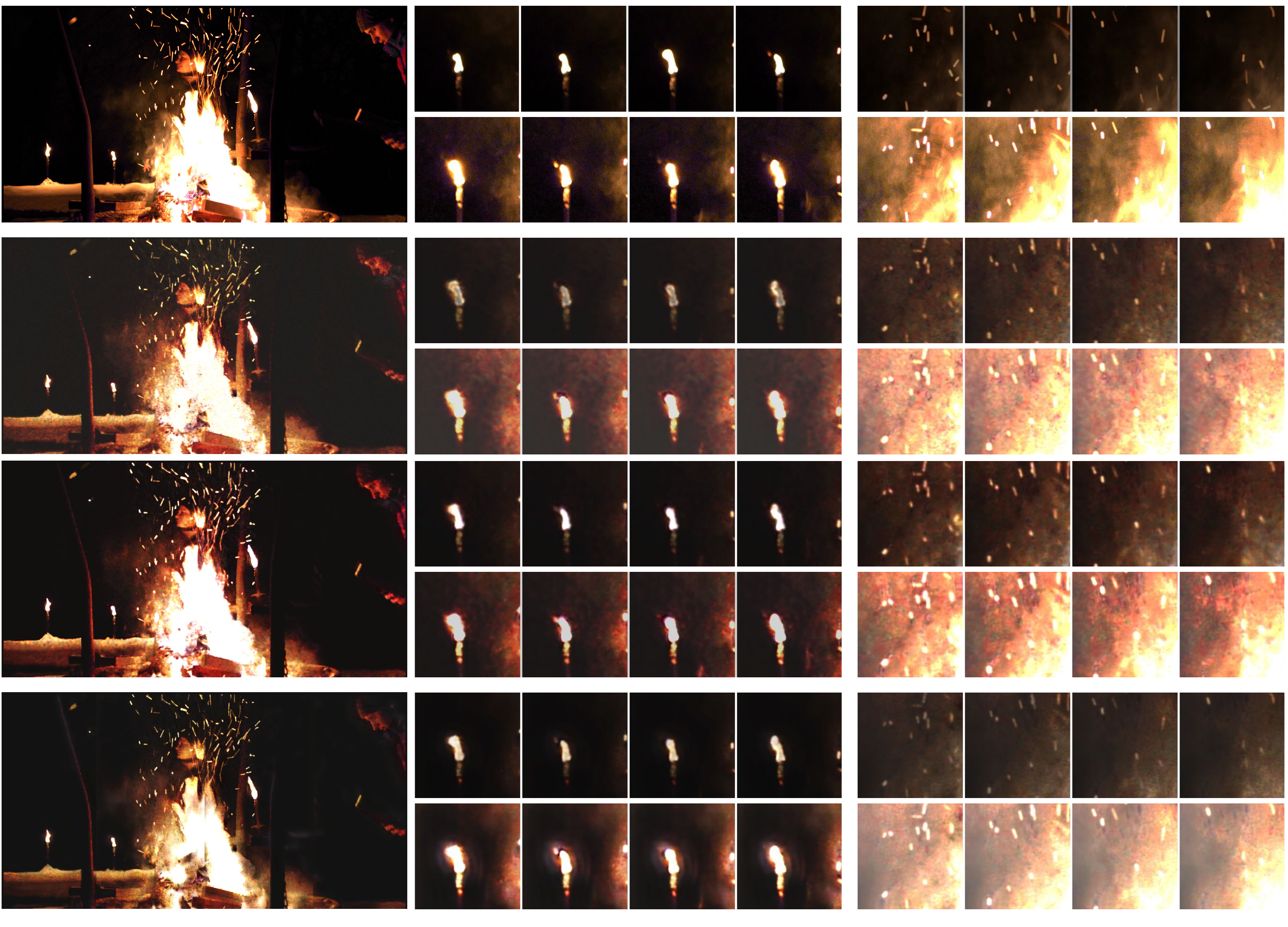}
	\put(4,73){full-resolution frame}
	\put(32,73){{consective frames~(zoom-in)}}
	\put(67,73){{consective frames~(zoom-in)}}
%		\put(79,90){proposed}
%		\put(21,1){\rotatebox{90}{\,\small {HDR-VDP-2}}}	
	
	\put(-2,60){\rotatebox{90}{\, {GT}}}
	\put(-2,40){\rotatebox{90}{\, {C3D}}}
	\put(-2,23){\rotatebox{90}{\, {DC3D}}}
	\put(-2,4){\rotatebox{90}{\,  {C3DED}}}
		\end{overpic}};
	%\begin{scope}[x={(image.south east)},y={(image.north west)}]
	%\draw[color=green, thick] (0.001,0.53) rectangle ++(0.055,0.17);
	%\draw[color=green, thick] (0.23,0.68) rectangle ++(0.055,0.17);
	%\draw[color=yellow, thick] (0.135,0.095) rectangle ++(0.055,0.17);
	%\draw[color=yellow, thick] (0.06,0.015) rectangle ++(0.055,0.17);
	%\end{scope}

	\end{tikzpicture}
	\caption[Simulation results for HDR video reconstruction.]{Simulated video HDR reconstruction results using the proposed 3D models of the $L_{VHDR}$ loss function. Left: one full-resolution frame tone-mapped from the recovered HDR video. Middle and right: zoom-in of two crops taken from %4 
	four consecutive frames, two recovered %exposure 
	exposures are shown for each crop.}
	\label{fig_ch6:sim_results_ab_1}
	\vspace{-2mm}
\end{figure}

\begin{figure}[hbt!]
	\centering
	\vspace{5 mm}
	\begin{tikzpicture}
	\node[anchor=south west,inner sep=0] (image) at (0,0) 
	{\begin{overpic}[width=1\columnwidth]{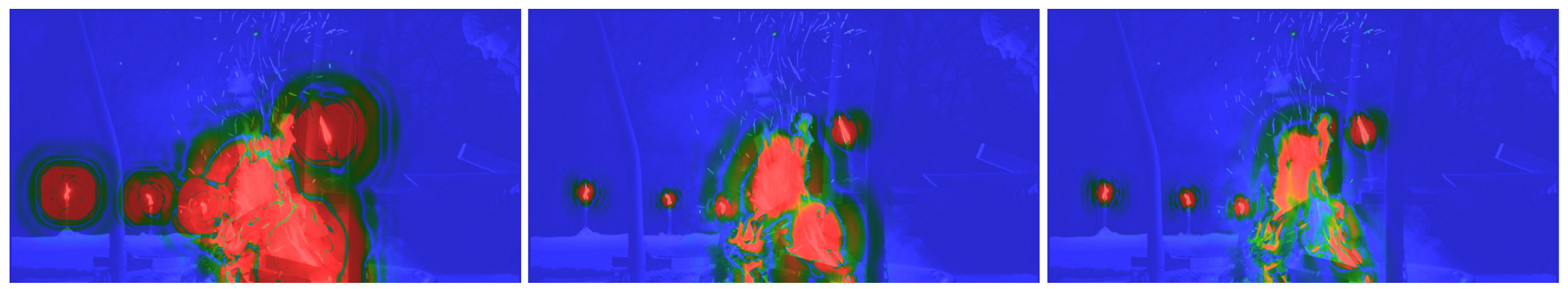}
		\put(13,19){C3D}
		\put(45,19){{DC3D}}
		\put(77,19){{C3DED}}
	
		\end{overpic}};

	\end{tikzpicture}
	\caption[%Simulation 
	Simulated HDR reconstruction comparison using HDR-VDP2.]{HDR-VDP2 results for the reconstructed frame shown in Figure~\ref{fig_ch6:sim_results_ab_1}, using~$L_{VHDR}$ loss. This fire scene is a challenging %30 stops 
	30-stop scene. The HDR-VDP2 maps indicate that the C3DED network results are the closest to %GT
	the results of GT.}
	\label{fig_ch6:Vid_vdp_res1}
	\vspace{-2mm}
\end{figure}
We synthesized 200 clips from the test shots as described in the previous section. The performance of each model with every loss function %measured
was measured using the average HDR-VDP2 $Q_{Score}$~\cite{mantiuk2011hdr}; larger values are better~(up to 100). %Also we compute 
We also computed the mean absolute error~(MAE), mean squared error~(MSE),  temporal loss~$L_T$~\eqref{eq:ch6_Tem_loss}, SSIM~\cite{wang2004image}, and perceptual loss~$L_{VGG}$~\eqref{eq5:perceptual_loss}. As indicated in the previous section, we %experiment 
experimented with $\lambda_2, \lambda_3$~, and $\lambda_4$ values. For each model, we trained ten instances, one for each the ten hyper-parameter values. %Results 
The results reported in this paper are for the hyper-parameters that produced the highest mean HDR-VDP2 $Q_{Score}$.  For all models, $L_{content}$~\eqref{eq:ch6_content_loss} with $\lambda_1=1, \lambda_2=0.01$ produced higher mean HDR-VDP2 $Q_{Score}$. Also $L_{VHDR}$~\eqref{eq:ch6_loss} for all models, $\lambda_1=1, \lambda_2=0.01, \lambda_3=\lambda_4=0.001$ led to the highest HDR-VDP mean score.  From Table~\ref{ch6:sim_res}, we can see that five out of the six metrics indicated that the C3DED network produced the best results, this could be related to the higher depth and receptive field of the C3DED network compared with the other two networks. Also, using $L_{VHDR}$ %produced in general 
generally produced better metric values for each model than $L_{Content}$. Although DC3D network have the same depth and number of parameters as the C3D networks, the numerical results for all %the 
metrics show that the DC3D network is better than the C3D network. This is most likely because of the increased receptive field resulted from utilizing the dilated 3D convolutions.

\begin{figure}[hbt!]
	\centering
	\vspace{3 mm}
	\begin{tikzpicture}
	\node[anchor=south west,inner sep=0] (image) at (0,0) 
	{\begin{overpic}[width=1\columnwidth]{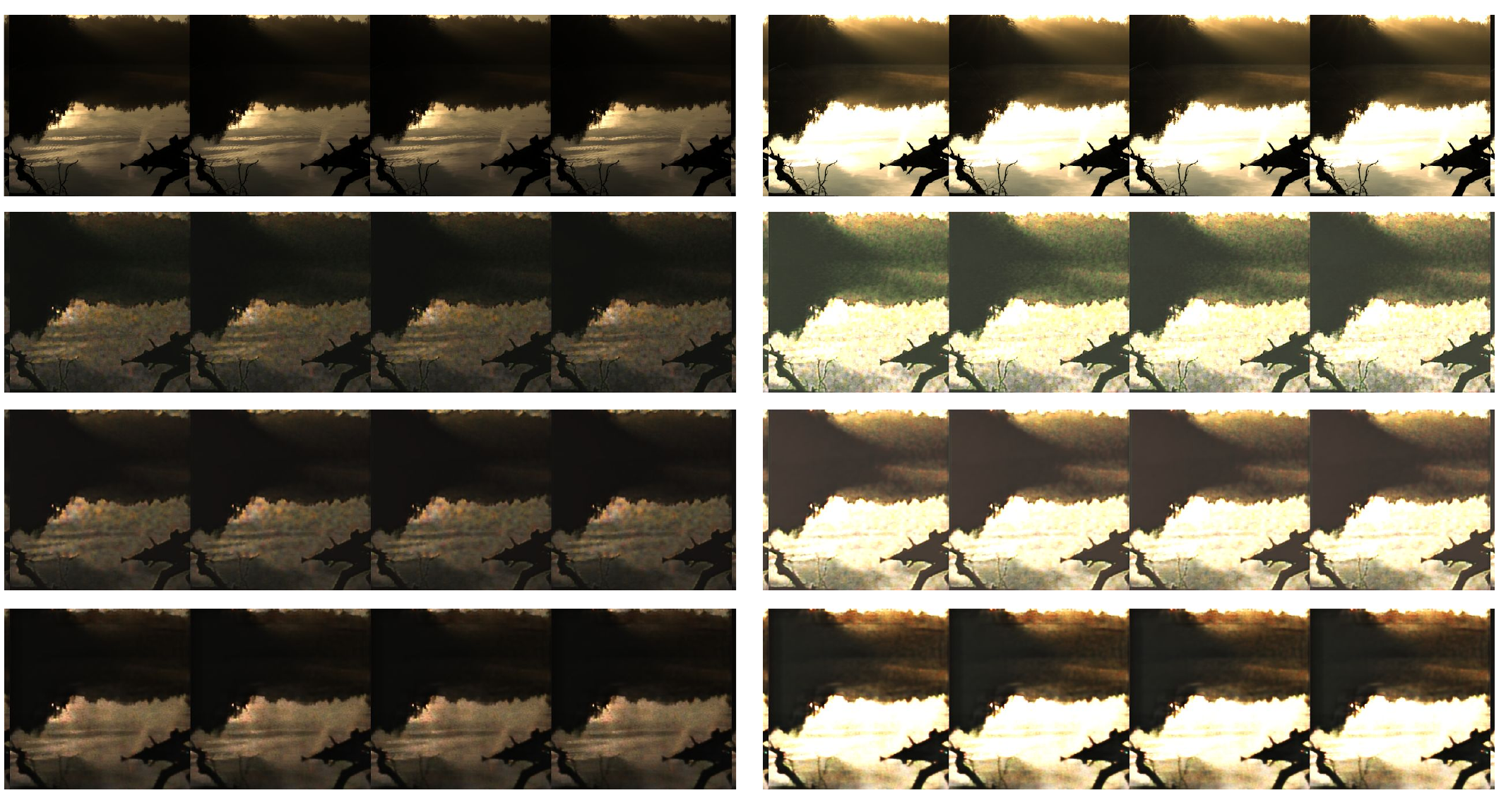}
		
		\put(7,55){{low (consective frames zoom-in)}}
		\put(56,55){{high (consective frames zoom-in)}}

		\put(-3,44){\rotatebox{90}{\, {GT}}}
		\put(-3,29){\rotatebox{90}{\, {C3D}}}
		\put(-3,16){\rotatebox{90}{\, {DC3D}}}
		\put(-3,1){\rotatebox{90}{\,  {C3DED}}}
		%		\put(9,90){GT}
		%		\put(28.5,90){\cite{nayar2000high}}
		%		\put(51,90){\cite{serrano2016convolutional}}
		%		\put(79,90){proposed}
		%		\put(21,1){\rotatebox{90}{\,\small {HDR-VDP-2}}}	
		\end{overpic}};
	%\begin{scope}[x={(image.south east)},y={(image.north west)}]
	%\draw[color=green, thick] (0.001,0.53) rectangle ++(0.055,0.17);
	%\draw[color=green, thick] (0.23,0.68) rectangle ++(0.055,0.17);
	%\draw[color=yellow, thick] (0.135,0.095) rectangle ++(0.055,0.17);
	%\draw[color=yellow, thick] (0.06,0.015) rectangle ++(0.055,0.17);
	%\end{scope}
	\end{tikzpicture}
	\caption[Zoom-in simulation results for HDR video reconstruction.]{Zoom-in simulated HDR video  reconstruction results using the proposed 3D models with the $L_{VHDR}$~loss function. The reconstructed tone-mapped HDR images with low exposure (left) and high exposure (right), indicating the  dynamic range of the recovered images. C3DED network images are the closest to the GT %ones
	images.}
	\label{fig_ch6:sim_results_ab_2}
	%\vspace{-2mm}
\end{figure}

\begin{figure}[hbt!]
	\centering
	\vspace{3 mm}
	\begin{tikzpicture}
	\node[anchor=south west,inner sep=0] (image) at (0,0) 
	{\begin{overpic}[width=1\columnwidth]{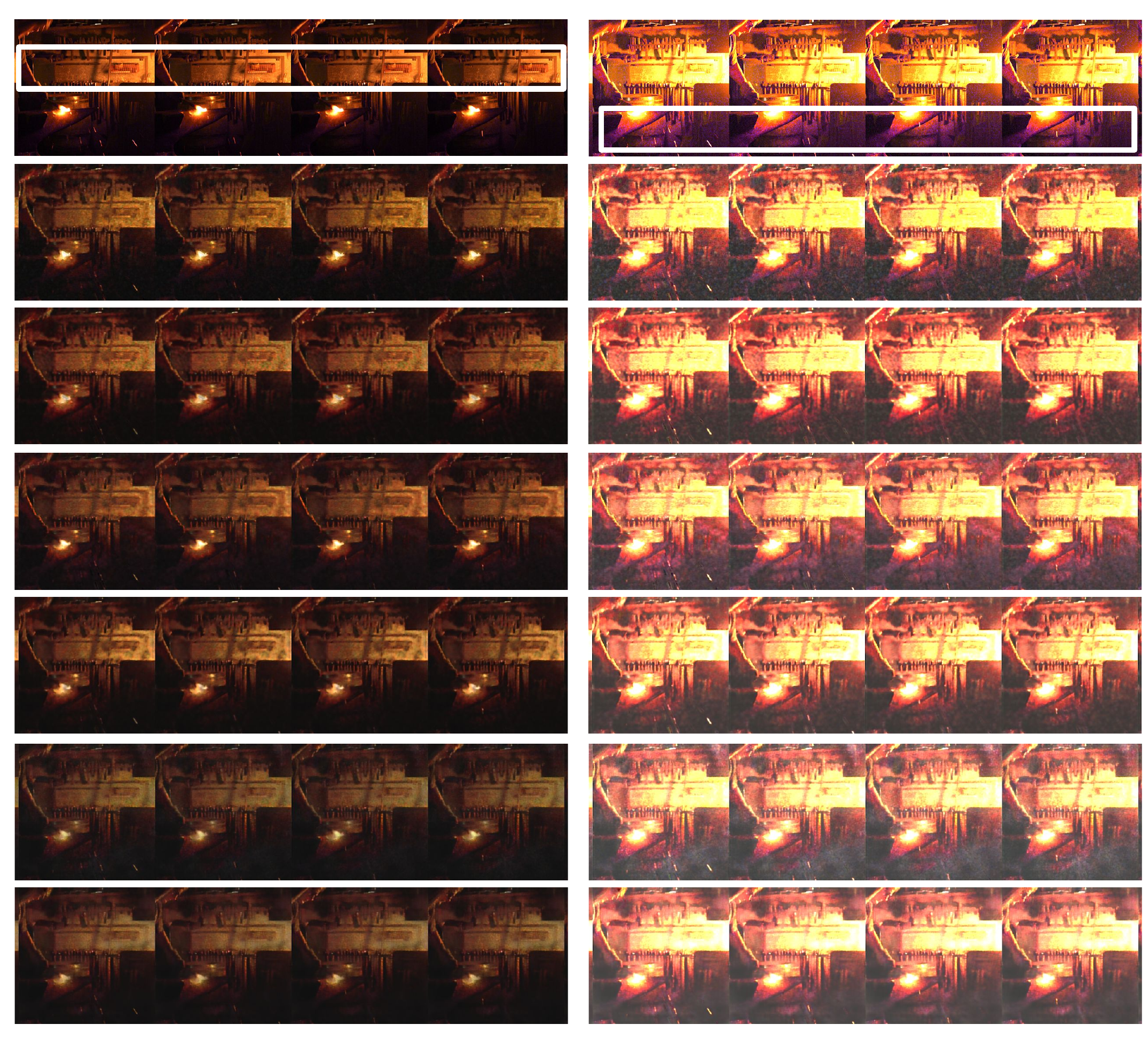}
		%		\put(9,90){GT}
		%		\put(28.5,90){\cite{nayar2000high}}
		%		\put(51,90){\cite{serrano2016convolutional}}
		%		\put(79,90){proposed}
		%		\put(21,1){\rotatebox{90}{\,\small {HDR-VDP-2}}}
			\put(7,90){{low consective frames~(zoom-in)}}
			\put(54,90){{high consective frames~(zoom-in)}}
			\put(-6,79){\rotatebox{90}{\, {GT}}}
			\put(-6,60){\rotatebox{90}{\, {C3D}}}
			\put(-6,33){\rotatebox{90}{\, {DC3D}}}
			\put(-6,7){\rotatebox{90}{\,  {C3DED}}}
			
			\put(-2,64){\rotatebox{90}{\, {$L_{Content}$}}}
			\put(-2,40){\rotatebox{90}{\, {$L_{Content}$}}}
			%\put(-2,22){\rotatebox{90}{\, {$L_{Content}$}}}
			\put(-2,13){\rotatebox{90}{\,  {$L_{Content}$}}}
			
			\put(-2,52){\rotatebox{90}{\, {$L_{VHDR}$}}}
			%\put(-2,40){\rotatebox{90}{\, {$L_{VHDR}$}}}
			\put(-2,27){\rotatebox{90}{\, {$L_{VHDR}$}}}
			\put(-2,1){\rotatebox{90}{\,  {$L_{VHDR}$}}}
		\end{overpic}};
	%\begin{scope}[x={(image.south east)},y={(image.north west)}]
	%\draw[color=green, thick] (0.001,0.53) rectangle ++(0.055,0.17);
	%\draw[color=green, thick] (0.23,0.68) rectangle ++(0.055,0.17);
	%\draw[color=yellow, thick] (0.135,0.095) rectangle ++(0.055,0.17);
	%\draw[color=yellow, thick] (0.06,0.015) rectangle ++(0.055,0.17);
	%\end{scope}
	
	\end{tikzpicture}
	\caption[Zoom-in simulation results for HDR video reconstruction.]{Zoom-in simulated HDR video reconstruction results using the proposed 3D models. Results for both $L_{VHDR}$ and $L_{content}$ loss function are shown. The reconstructed (zoom-in) tone-mapped HDR images for both low (left) and high (right) exposures %are indicating 
	indicate the dynamic range of the recovered results. Noise in the arias noted by the white box in the GT crops is reduced by the networks trained with $L_{VHDR}$ loss.}
	\label{fig_ch6:sim_results_ab_3}
	\vspace{-2mm}
\end{figure}

%Figure~\ref{fig_ch6:sim_results_ab_1} shows an example of a challenging 30 stops fire scene at night,  we present HDR reconstructed video results for the three networks with $L_{VHDR}$ loss. The two zoom-in crops show the details in four consecutive frames of the reconstructed clip. For each crop, we present two exposures indicating the recovered results' dynamic range. The three models can successfully construct a reliable HDR image. Figure~\ref{fig_ch6:Vid_vdp_res1} shows the HDR-VDP2 maps for the reconstructed frames in 

An example of a challenging %30 stops fire
30-stop fire scene at night using $L_{VHDR}$ loss is shown in Figure~\ref{fig_ch6:sim_results_ab_1}. The two zoom-in crops show the details in four consecutive frames of the reconstructed clip. For each crop, we present two exposures indicating %the recovered results' dynamic range
the high dynamic range of the recovered results. The three models can successfully construct a reliable HDR image. Figure~\ref{fig_ch6:Vid_vdp_res1} shows the HDR-VDP2 maps each full-resolution frame shown in Figure~\ref{fig_ch6:sim_results_ab_1}. The HDR-VDP2 maps indicate that the %C3DED network results 
results of the C3DED network are the closest to the ground truth HDR image. Figure~\ref{fig_ch6:sim_results_ab_2} shows a zoom-in %for 
of another challenging scene. All results in this figure %obtained 
were obtained 
using the network trained with~$L_{VHDR}$ loss. %C3DED results 
The results of the C3DED network are the closest to ground truth HDR image.

Figure~\ref{fig_ch6:sim_results_ab_3} shows the effect of using $L_{VHDR}$ instead of $L_{content}$ %in 
on noise and artifact reduction. This figure shows zoom-in images of HDR video reconstructed frames using the proposed 3D models. Results for both the $L_{VHDR}$ and $L_{content}$ loss function are shown. Noise in the arias noted by the white box in GT crops is reduced by the networks trained using $L_{VHDR}$ loss.

\begin{figure}[hbt!]
	\centering
	\vspace{3 mm}
	\begin{tikzpicture}
	\node[anchor=south west,inner sep=0] (image) at (0,0) 
	{\begin{overpic}[width=.6\columnwidth]{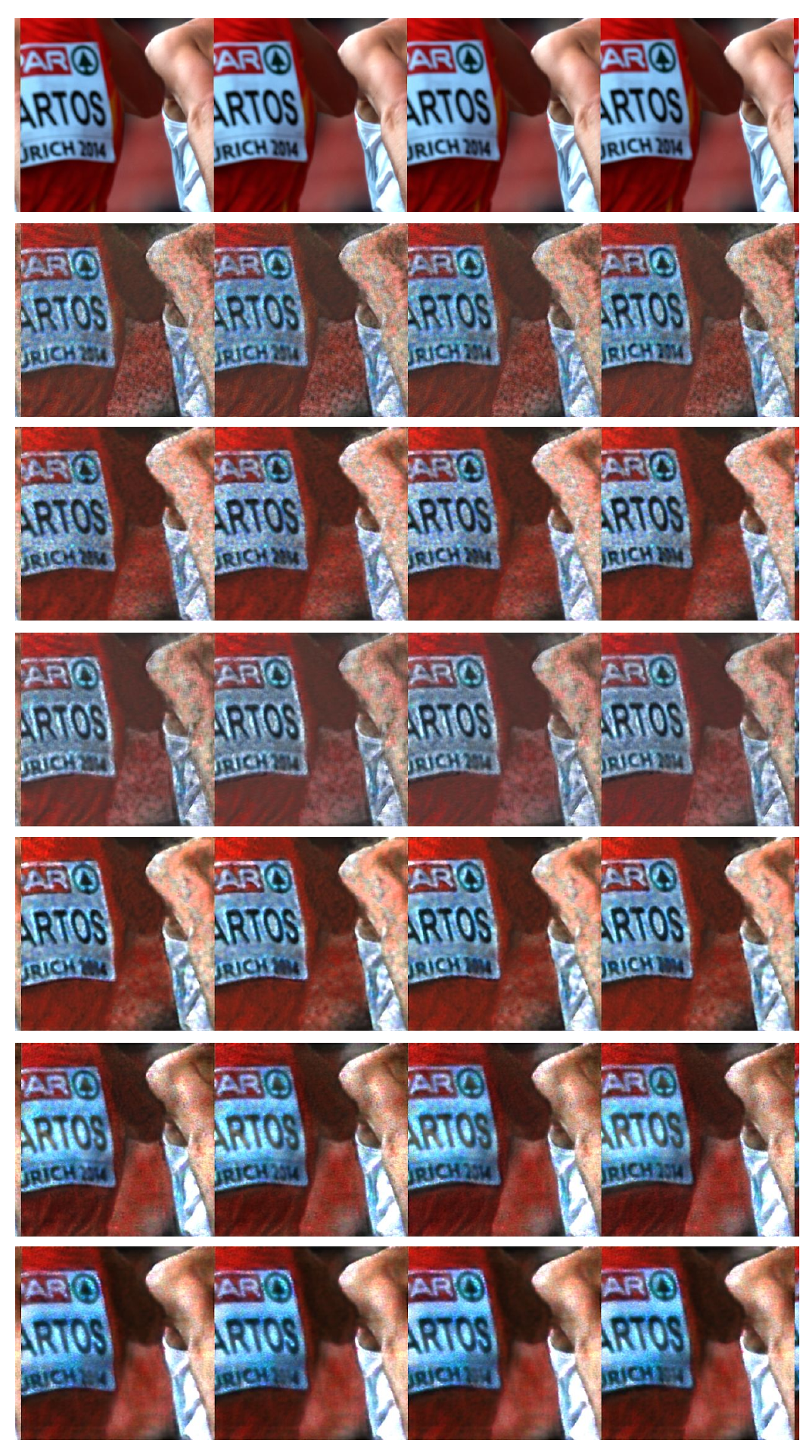}
		\put(9,100){{four consecutive frames~(zoom-in)}}
		\put(-6,88){\rotatebox{90}{\, {GT}}}
		\put(-6,67){\rotatebox{90}{\, {C3D}}}
		\put(-6,38){\rotatebox{90}{\, {DC3D}}}
		\put(-6,9){\rotatebox{90}{\,  {C3DED}}}
		
		\put(-2,72){\rotatebox{90}{\, {$L_{Content}$}}}
		\put(-2,45){\rotatebox{90}{\, {$L_{Content}$}}}
		%\put(-2,22){\rotatebox{90}{\, {$L_{Content}$}}}
		\put(-2,16){\rotatebox{90}{\,  {$L_{Content}$}}}
		
		\put(-2,59){\rotatebox{90}{\, {$L_{VHDR}$}}}
		%\put(-2,40){\rotatebox{90}{\, {$L_{VHDR}$}}}
		\put(-2,31){\rotatebox{90}{\, {$L_{VHDR}$}}}
		\put(-2,3){\rotatebox{90}{\,  {$L_{VHDR}$}}}	
		\end{overpic}};
	%\begin{scope}[x={(image.south east)},y={(image.north west)}]
	%\draw[color=green, thick] (0.001,0.53) rectangle ++(0.055,0.17);
	%\draw[color=green, thick] (0.23,0.68) rectangle ++(0.055,0.17);
	%\draw[color=yellow, thick] (0.135,0.095) rectangle ++(0.055,0.17);
	%\draw[color=yellow, thick] (0.06,0.015) rectangle ++(0.055,0.17);
	%\end{scope}
	\end{tikzpicture}
	\caption[The effect of using $L_{VHDR}$ %for %results 
	on temporal coherency.]{Zoom-in simulated HDR video reconstruction results using the proposed 3D models. Both $L_{VHDR}$ and $L_{content}$ loss function recovered images are shown. Using $L_{VHDR}$ generates more temporally coherent results for the three models. The C3DED model recovered images have much less noise than the other two models.}
	\label{fig_ch6:sim_results_ab_4}
	\vspace{-2mm}
\end{figure}

Figure~\ref{fig_ch6:sim_results_ab_4} shows the effect of using $L_{VHDR}$ instead of $L_{content}$ %in
on generation of more temporally coherent results. This figure shows zoom-in of HDR video frames reconstructed using the proposed 3D models. For all networks, using~$L_{VHDR}$ results in a more %temporal 
temporally coherent image. This scene is a challenging running clip simulated from the Zurich Athletics 2014 dataset~\cite{EBU_2014}. We can see both the C3D and DC3D results are degraded for this test clip, indicating that such scenes need a higher network depth and receptive field.

\section{Discussion}
This paper shows how to utilize 3D convolutional networks to recover HDR video from coded LDR video. We experiment with %a 
temporal loss. The obtained results are promising and could lead to affordable HDR video %capturing 
capture using conventional cameras. However, several aspects need further investigation. In particular, as mentioned in the training section, we used 3K simulated coded clips (24K coded CFA LDR frames in total) from the 45 training shots (59K HDR frames in total) in the training dataset. In general, deep learning networks need to be trained on a much larger dataset to produce more accurate results%,
. This raises two questions: will generating more simulated clips from HDR training shots be enough? Will pre-training on a larger synthesized dataset from high-quality LDR videos, as we performed in ~\cite{alghamdi2019}, it was necessary to pre-train the inception network to reconstruct still HDR images from CFA coded LDR? 

In \cite{alghamdi2021Transfer}, we addressed this issue by transfer learning from a different large-scale task (image classification on ImageNet), %leading 
which led to considerable improvements in still HDR reconstruction. However, we tried a similar approach using a residual 3D encoder-decoder network for video reconstruction. We %test
tested an encoder network pre-trained for action recognition on Kinetics-400~\cite{kay2017kinetics}. Specifically, we %select 
selected R3D-18, which is an 18 layer Resnet3D network~\cite{tran2018closer}. We trained this network %with 
on both 3K training samples and 30K training samples. For both datasets, the network suffers from overfitting to the training data. This overfitting could be solved using a larger training set.  However, another downside of this approach is the considerable memory requirement for such a network.

\bibliographystyle{unsrtnat}
\bibliography{references}  %%% Uncomment this line and comment out the ``thebibliography'' section below to use the external .bib file (using bibtex) .

\begin{thebibliography}{37}
\providecommand{\natexlab}[1]{#1}
\providecommand{\url}[1]{\texttt{#1}}
\expandafter\ifx\csname urlstyle\endcsname\relax
  \providecommand{\doi}[1]{doi: #1}\else
  \providecommand{\doi}{doi: \begingroup \urlstyle{rm}\Url}\fi

\bibitem[Banterle et~al.(2017)Banterle, Artusi, Debattista, and
  Chalmers]{artusi2017advanced}
Francesco Banterle, Alessandro Artusi, Kurt Debattista, and Alan Chalmers.
\newblock \emph{Advanced high dynamic range imaging}.
\newblock AK Peters/CRC Press, New York, 2017.
\newblock ISBN 9781315119526.
\newblock \doi{10.1201/9781315119526}.

\bibitem[Alghamdi et~al.(2019)Alghamdi, Fu, Thabet, and Heidrich]{alghamdi2019}
Masheal Alghamdi, Qiang Fu, Ali Thabet, and Wolfgang Heidrich.
\newblock {Reconfigurable Snapshot HDR Imaging Using Coded Masks and Inception
  Network}.
\newblock In Hans-Jörg Schulz, Matthias Teschner, and Michael Wimmer, editors,
  \emph{Vision, Modeling and Visualization}. The Eurographics Association,
  2019.
\newblock ISBN 978-3-03868-098-7.
\newblock \doi{10.2312/vmv.20191316}.

\bibitem[Alghamdi et~al.(2021)Alghamdi, Fu, Thabet, and
  Heidrich]{alghamdi2021Transfer}
Masheal Alghamdi, Qiang Fu, Ali Thabet, and Wolfgang Heidrich.
\newblock Transfer deep learning for reconfigurable snapshot hdr imaging using
  coded masks.
\newblock \emph{Computer Graphics Forum}, 2021.
\newblock \doi{https://doi.org/10.1111/cgf.14205}.
\newblock URL \url{https://onlinelibrary.wiley.com/doi/abs/10.1111/cgf.14205}.

\bibitem[Nayar and Mitsunaga(2000)]{nayar2000high}
Shree~K Nayar and Tomoo Mitsunaga.
\newblock High dynamic range imaging: {S}patially varying pixel exposures.
\newblock In \emph{IEEE Conference on Computer Vision and Pattern Recognition
  (CVPR)}, volume~1, pages 472--479. IEEE, 2000.

\bibitem[Nayar and Branzoi(2003)]{nayar2003adaptive}
Shree~K Nayar and Vlad Branzoi.
\newblock Adaptive dynamic range imaging: Optical control of pixel exposures
  over space and time.
\newblock In \emph{IEEE International Conference on Computer Vision (ICCV)},
  volume~2, pages 1168--1175. IEEE, 2003.

\bibitem[Tocci et~al.(2011)Tocci, Kiser, Tocci, and Sen]{tocci2011versatile}
Michael~D. Tocci, Chris Kiser, Nora Tocci, and Pradeep Sen.
\newblock A versatile {HDR} video production system.
\newblock In \emph{ACM SIGGRAPH 2011}, pages 41:1--41:10, New York, NY, USA,
  2011. ACM.
\newblock ISBN 978-1-4503-0943-1.

\bibitem[Chalmers and Debattista(2011)]{chalmers2011hdr}
Alan Chalmers and Kurt Debattista.
\newblock Hdr video: Capturing and displaying dynamic real-world lighting.
\newblock In \emph{Color and Imaging Conference}, volume 2011, pages 177--180.
  Society for Imaging Science and Technology, 2011.

\bibitem[{Red Company}()]{RedCompany}
{Red Company}.
\newblock Red one.
\newblock URL \url{https://www.red.com/}.

\bibitem[{Grass Valley}()]{GrassValley}
{Grass Valley}.
\newblock Thomson grass valley.
\newblock URL \url{https://www.grassvalley.com/products/}.

\bibitem[{Arri Alexa}()]{ArriAlexa}
{Arri Alexa}.
\newblock Alexa cameras.
\newblock URL \url{https://www.arri.com/en/camera-systems/cameras}.

\bibitem[Sony()]{Sony}
Sony.
\newblock Sony pxw-z90.
\newblock URL
  \url{https://pro.sony/ue{\_}US/products/handheld-camcorders/broadcast-palm-sized-4k-camcorders-everyone}.

\bibitem[{Vision Research}()]{VisionResearch}
{Vision Research}.
\newblock Phantom hd.
\newblock URL \url{http://www.visionresearch.com/}.

\bibitem[Reinhard et~al.(2010)Reinhard, Heidrich, Debevec, Pattanaik, Ward, and
  Myszkowski]{reinhard2010high}
Erik Reinhard, Wolfgang Heidrich, Paul Debevec, Sumanta Pattanaik, Greg Ward,
  and Karol Myszkowski.
\newblock \emph{High dynamic range imaging: acquisition, display, and
  image-based lighting}.
\newblock Morgan Kaufmann, 2010.

\bibitem[Debevec and Malik(1997)]{debevec1997recovering}
Paul~E Debevec and Jitendra Malik.
\newblock Recovering high dynamic range radiance maps from photographs.
\newblock In \emph{Proceedings of the 24th annual conference on Computer
  graphics and interactive techniques}, pages 369--378. ACM
  Press/Addison-Wesley Publishing Co., 1997.

\bibitem[Mann et~al.(1995)Mann, Picard, Mann, and Picard]{Picard1995onbeing}
Mann, Picard, S.~Mann, and R.~W. Picard.
\newblock On being ``undigital'' with digital cameras: extending dynamic range
  by combining differently combining differently exposed pictures.
\newblock In \emph{Proceedings of IS\&T 1995}, pages 442--448, 1995.

\bibitem[McGuire et~al.(2007)McGuire, Matusik, Pfister, Chen, Hughes, and
  Nayar]{mcguire2007optical}
Morgan McGuire, Wojciech Matusik, Hanspeter Pfister, Billy Chen, John~F Hughes,
  and Shree~K Nayar.
\newblock Optical splitting trees for high-precision monocular imaging.
\newblock \emph{IEEE Computer Graphics and Applications}, 27\penalty0 (2),
  2007.

\bibitem[Kronander et~al.(2013)Kronander, Gustavson, Bonnet, and
  Unger]{kronander2013unified}
Joel Kronander, Stefan Gustavson, Gerhard Bonnet, and Jonas Unger.
\newblock Unified {HDR} reconstruction from raw {CFA} data.
\newblock In \emph{IEEE International Conference on Computational Photography
  (ICCP) 2013}, pages 1--9. IEEE, 2013.

\bibitem[Serrano et~al.(2016)Serrano, Heide, Gutierrez, Wetzstein, and
  Masia]{serrano2016convolutional}
Ana Serrano, Felix Heide, Diego Gutierrez, Gordon Wetzstein, and Belen Masia.
\newblock Convolutional sparse coding for high dynamic range imaging.
\newblock In \emph{Computer Graphics Forum}, volume~35, pages 153--163. Wiley
  Online Library, 2016.

\bibitem[Kensei et~al.(2014)Kensei, Kaizu, and Mitsunaga]{kensei2014image}
JO~Kensei, Shun Kaizu, and Tomoo Mitsunaga.
\newblock Image processing including image correction, September~30 2014.
\newblock US Patent 8,848,063.

\bibitem[Ronneberger et~al.(2015)Ronneberger, Fischer, and
  Brox]{ronneberger2015u}
Olaf Ronneberger, Philipp Fischer, and Thomas Brox.
\newblock U-net: Convolutional networks for biomedical image segmentation.
\newblock In \emph{International Conference on Medical image computing and
  computer-assisted intervention}, pages 234--241. Springer, 2015.

\bibitem[Simonyan and Zisserman(2014)]{simonyan2014very}
Karen Simonyan and Andrew Zisserman.
\newblock Very deep convolutional networks for large-scale image recognition.
\newblock \emph{arXiv preprint}, 2014.

\bibitem[Ji et~al.(2012)Ji, Xu, Yang, and Yu]{ji20123d}
Shuiwang Ji, Wei Xu, Ming Yang, and Kai Yu.
\newblock 3d convolutional neural networks for human action recognition.
\newblock \emph{IEEE transactions on pattern analysis and machine
  intelligence}, 35\penalty0 (1):\penalty0 221--231, 2012.

\bibitem[Tran et~al.(2015)Tran, Bourdev, Fergus, Torresani, and
  Paluri]{tran2015learning}
Du~Tran, Lubomir Bourdev, Rob Fergus, Lorenzo Torresani, and Manohar Paluri.
\newblock Learning spatiotemporal features with 3d convolutional networks.
\newblock In \emph{Proceedings of the IEEE international conference on computer
  vision}, pages 4489--4497, 2015.

\bibitem[Lv et~al.(2018)Lv, Lu, Wu, and Lim]{lv2018mbllen}
Feifan Lv, Feng Lu, Jianhua Wu, and Chongsoon Lim.
\newblock Mbllen: Low-light image/video enhancement using cnns.
\newblock In \emph{BMVC}, page 220, 2018.

\bibitem[{Kappeler} et~al.(2016){Kappeler}, {Yoo}, {Dai}, and
  {Katsaggelos}]{Kappeler2016VideoSuper}
A.~{Kappeler}, S.~{Yoo}, Q.~{Dai}, and A.~K. {Katsaggelos}.
\newblock Video super-resolution with convolutional neural networks.
\newblock \emph{IEEE Transactions on Computational Imaging}, 2\penalty0
  (2):\penalty0 109--122, 2016.
\newblock \doi{10.1109/TCI.2016.2532323}.

\bibitem[Wang et~al.(2017)Wang, Huang, You, Yang, and Neumann]{Wang_2017_ICCV}
Weiyue Wang, Qiangui Huang, Suya You, Chao Yang, and Ulrich Neumann.
\newblock Shape inpainting using 3d generative adversarial network and
  recurrent convolutional networks.
\newblock In \emph{Proceedings of the IEEE International Conference on Computer
  Vision (ICCV)}, Oct 2017.

\bibitem[Wan(2019)]{Wang_Huang_Han_Wang_2019}
Video inpainting by jointly learning temporal structure and spatial details.
\newblock 33:\penalty0 5232--5239, Jul. 2019.
\newblock \doi{10.1609/aaai.v33i01.33015232}.
\newblock URL \url{https://ojs.aaai.org/index.php/AAAI/article/view/4458}.

\bibitem[Konnik and Welsh(2014)]{konnik2014high}
Mikhail Konnik and James Welsh.
\newblock High-level numerical simulations of noise in {CCD} and {CMOS}
  photosensors: review and tutorial.
\newblock \emph{arXiv preprint}, 2014.

\bibitem[Yu and Koltun(2015)]{yu2015multi}
Fisher Yu and Vladlen Koltun.
\newblock Multi-scale context aggregation by dilated convolutions.
\newblock \emph{arXiv preprint arXiv:1511.07122}, 2015.

\bibitem[Banitalebi-Dehkordi et~al.(2014)Banitalebi-Dehkordi, Azimi, Pourazad,
  and Nasiopoulos]{banitalebi2014compression}
Amin Banitalebi-Dehkordi, Maryam Azimi, Mahsa~T Pourazad, and Panos
  Nasiopoulos.
\newblock Compression of high dynamic range video using the {HEVC} and
  {H.264/AVC} standards.
\newblock In \emph{10th International Conference on Heterogeneous Networking
  for Quality, Reliability, Security and Robustness}, pages 8--12. IEEE, 2014.

\bibitem[EBU(2014)]{EBU_2014}
EBU.
\newblock Zurich athletics 2014 dataset.
\newblock \url{https://tech.ebu.ch/testsequences/zurich_athletics}, 2014.

\bibitem[Paszke et~al.(2017)Paszke, Gross, Chintala, Chanan, Yang, DeVito, Lin,
  Desmaison, Antiga, and Lerer]{paszke2017automatic}
Adam Paszke, Sam Gross, Soumith Chintala, Gregory Chanan, Edward Yang, Zachary
  DeVito, Zeming Lin, Alban Desmaison, Luca Antiga, and Adam Lerer.
\newblock Automatic differentiation in pytorch.
\newblock In \emph{NIPS-W}, 2017.

\bibitem[Kingma and Ba(2015)]{kingma2014adam}
Diederik~P. Kingma and Jimmy Ba.
\newblock Adam: {A} method for stochastic optimization.
\newblock In Yoshua Bengio and Yann LeCun, editors, \emph{3rd International
  Conference on Learning Representations, {ICLR} 2015, San Diego, CA, USA, May
  7-9, 2015, Conference Track Proceedings}, 2015.
\newblock URL \url{http://arxiv.org/abs/1412.6980}.

\bibitem[Mantiuk et~al.(2011)Mantiuk, Kim, Rempel, and
  Heidrich]{mantiuk2011hdr}
Rafat Mantiuk, Kil~Joong Kim, Allan~G Rempel, and Wolfgang Heidrich.
\newblock {HDR-VDP-2}: {A} calibrated visual metric for visibility and quality
  predictions in all luminance conditions.
\newblock In \emph{ACM Trans. Graph.}, volume~30, page~40. ACM, 2011.

\bibitem[Wang et~al.(2004)Wang, Bovik, Sheikh, Simoncelli,
  et~al.]{wang2004image}
Zhou Wang, Alan~C Bovik, Hamid~R Sheikh, Eero~P Simoncelli, et~al.
\newblock Image quality assessment: from error visibility to structural
  similarity.
\newblock \emph{IEEE transactions on image processing}, 13\penalty0
  (4):\penalty0 600--612, 2004.

\bibitem[Kay et~al.(2017)Kay, Carreira, Simonyan, Zhang, Hillier,
  Vijayanarasimhan, Viola, Green, Back, Natsev, Suleyman, and
  Zisserman]{kay2017kinetics}
Will Kay, Joao Carreira, Karen Simonyan, Brian Zhang, Chloe Hillier, Sudheendra
  Vijayanarasimhan, Fabio Viola, Tim Green, Trevor Back, Paul Natsev, Mustafa
  Suleyman, and Andrew Zisserman.
\newblock The kinetics human action video dataset, 2017.

\bibitem[Tran et~al.(2018)Tran, Wang, Torresani, Ray, LeCun, and
  Paluri]{tran2018closer}
Du~Tran, Heng Wang, Lorenzo Torresani, Jamie Ray, Yann LeCun, and Manohar
  Paluri.
\newblock A closer look at spatiotemporal convolutions for action recognition,
  2018.

\end{thebibliography}





% \end{thebibliography}

\end{document}